\documentclass[twocolumn]{emulateapj}

\usepackage{graphicx}

\shorttitle{GRAVITY DARKENING AND BRIGHTENING IN BINARIES}
\shortauthors{WHITE, BAUMGARTE \& SHAPIRO}

\begin{document}
\title{Gravity darkening and brightening in binaries}
\author{Helen E.~White\altaffilmark{1},
Thomas W.~Baumgarte \altaffilmark{1,2} and
Stuart L.~Shapiro \altaffilmark{2,3}}
%
\affil
{\altaffilmark{1} 
Department of Physics and Astronomy, Bowdoin College,
Brunswick, ME 04011}
\affil
{\altaffilmark{2} 
Department of Physics, University of Illinois at Urbana-Champaign,
Urbana, IL 61801}
\affil
{\altaffilmark{3} 
Department of Astronomy and NCSA, University of Illinois at
Urbana-Champaign, Urbana, IL 61801}
%
\begin{abstract}
%
We apply a von Zeipel gravity darkening model to corotating binaries to obtain  a simple, analytical expression for the emergent radiative flux from a 
tidally distorted primary orbiting a point-mass secondary.
We adopt a simple Roche model to determine the envelope structure of 
the primary, assumed massive and centrally condensed, and use the results to
calculate the flux.  As for single rotating stars, 
gravity {\em darkening} reduces the flux along the stellar equator of the
primary, but, unlike for rotating stars, we find that gravity {\em brightening} enhances the flux in a region around the stellar poles.  We identify a critical limiting separation beyond which hydrostatic equilibrium no longer 
is possible, whereby the flux vanishes at the point on the stellar equator 
of the primary facing the companion.  For equal-mass binaries, the total luminosity is reduced by about 13\% when this limiting separation is reached.
\end{abstract}


%
%

\section{Introduction}

Early studies \citep{Zei24a,Zei24b,Zei24c,Cha33} showed that on stellar surfaces the radiative flux is proportional to the effective gravitational force 
\cite[see, e.g.,][for reviews]{KipW90,Tas00}.   In rotating stars, this means that regions close to the pole are brighter (and have a higher effective temperature) than regions close to the equator, an effect that is often referred to as {\em gravity darkening}.  In fact, for stars rotating at the break-up speed, the radiative flux vanishes at the equator \citep[see, e.g.,][hereafter Paper I]{BauS99}.   Gravity darkening plays an 
important role in the classification of stars \cite[e.g.][]{MaeP70}, and may even provide a means of estimating stellar masses independently of binary companions \citep{Zhaetal09}.

While the simplest gravity darkening models
make several assumptions that restrict their applicability to very
massive and supermassive stars 
(as we will review in Section \ref{sec:assumptions} below), they continue to be useful in a number of different contexts.  
\cite{CraO95}, for example, adopt similar models to compute radiative fluxes 
and driving forces for winds from rapidly rotating B stars. 
In Paper I, two of us 
adopted a simple Roche model to determine the envelope structure of and 
emergent radiation flux from a rotating, 
supermassive star (SMS) and employed the results to calculate the 
evolutionary timescale for these objects.

An exciting recent development provides a new motivation for studying
gravity darkening.  With optical or infrared interferometric arrays
(including the PTI, NPOI and CHARA arrays), it has become possible to
obtain resolved interferometric images of individual stars
\cite[e.g.][]{Beletal01,Petetal06,Aufetal06}.  Observations of
rotating stars, which indeed show oblate shapes with brighter poles
and darker equators, are analyzed using gravity darkening models.
These models can be used to relate different stellar features to each
other, for example the ratio between the polar and equatorial fluxes
to the star's angular velocity.  While some deviations between
observations and models point to the fact that some of the assumptions
made in the models may be too restrictive (for example, the observed
stars may have complicated surface layers or differential rotation;
see, e.g., \cite{Monetal07} as well as Section 7 of \cite{Zhaetal09}
for discussions), the basic features are represented reasonably well.
Moreover, the quality of fits can be improved by introducing
additional free parameters into the models (for other model
improvements see, e.g., \cite{EspR11} and \cite{Cla12}).

In addition to observations of rotating stars, resolved interferometric images of close binary stars have become available \citep{Zhaetal08}.  While these observations do not yet have sufficient resolution to distinguish detailed models, they do motivate a study of gravity darkening in binaries.  

In this short paper we point out that, under the assumptions listed in
Section \ref{sec:assumptions}, the simple gravity darkening models for
rotating stars can be generalized very easily to describe corotating
stars in a binary system.  As a consequence, we obtain very simple
analytical expressions for the envelope structure, flux and the total
luminosity for a massive or supermassive primary orbiting a point-mass
secondary as a function of the binary separation and mass ratio.
While the effects of rotation always decrease the surface flux
(compared to a nonrotating star in isolation), the presence of a
binary companion decreases the flux in some regions of the star, but
increases the flux in a region around the pole.  In a binary, we
therefore encounter both gravity darkening and gravity brightening.
We also identify a critical limiting separation below which
hydrostatic equilibrium cannot exist and at which the surface flux
from the point on the stellar equator facing the binary companion
vanishes.  For an equal-mass binary, the total luminosity is reduced
by about 13\% once this separation has been reached.

Elements of this problem have been known for a long time \citep[e.g.][]{Zei24c,Osa65} and the model we construct presumably 
will be superseded by more detailed numerical treatments.
We nevertheless hope that our simple analysis and analytical 
expressions may prove useful, for instance, in providing preliminary interpretations 
of future resolved images of close binary stars and calibrating more detailed numerical models.

Our paper is organized as follows.  In Section \ref{sec:assumptions} we review our basic assumptions.  In Section \ref{sec:roche} we review the Roche approximation and its application to
binaries to determine the envelope structure and locate the surface of corotating stars
in a binary.  In Section \ref{sec:flux} we compute the radiative flux from these stars, 
and find both gravity darkening and brightening.  We integrate this flux to find the total luminosity in Section \ref{sec:luminosity} and conclude with a brief discussion 
in Section \ref{sec:discussion}.

%
 
\section{Basic Assumptions}
\label{sec:assumptions}

Our analysis of a binary relies on several explicit assumptions.  Some of these assumptions we already adopted in Paper I, where we treated uniformly rotating, highly massive stars and SMSs in isolation.  To determine the equilibrium structure of the envelope, we assume that the primary is  

\begin{enumerate}

\item governed by Newtonian gravitation, 

\item in synchronous orbit about a point-mass secondary,

\end{enumerate}

and that its envelope

\begin{enumerate}
\setcounter{enumi}{2}

\item is described by the Roche model,

\item is characterized by a polytropic equation of state, and

\item interacts with the secondary via a potential that can be truncated beyond the
quadrupolar tidal term.

\end{enumerate}

To determine the emergent radiative flux from the primary, we further assume that its envelope is

\begin{enumerate}
\setcounter{enumi}{5}

\item dominated by thermal radiation pressure,

\item fully convective, and

\item characterized by a constant Rosseland mean opacity (e.g., electron scattering). 
\end{enumerate}

We assume that gravitational fields are sufficiently weak so that we
can apply Newtonian gravity.  This assumption clearly holds for normal stars. SMSs of greatest astrophysical interest have masses and radii 
that satisfy $R/M \gtrsim 400$  so that this assumption certainly holds.
Relativistic corrections are important for the stability of SMSs, but
can be neglected in the analysis of the equilibrium state.

The assumption that the binary is in corotation follows from
the observation that most short period binaries containing massive
stars orbit in circular, synchronous orbits \citep{Vanetal98}. 
The combination of convection and magnetic fields are likely to generate
an effective turbulent viscosity, which dampens nonsynchronous motion and
brings the binary into corotation.

The assumption of a point-mass companion is adopted for simplicity.
If the corotating companion is also a massive star, the point-mass can be
replaced by a finite star whose envelope can be treated identically to
the envelope of the primary. The results 
for the companion's envelope structure and emergent flux would depend 
the same way on its mass and radius as we find for the primary. 
We truncate the potential of the secondary after the quadrupolar tidal term, which captures the leading-order, dominant effects of the secondary on the primary.   

For large masses, the ratio between radiation pressure, $P_r$, and gas
pressure, $P_g$, satisfies
\begin{equation} \label{beta}
\beta_P \equiv \frac{P_g}{P_r} = 8.49 
        \left( \frac{M}{M_{\odot}} \right)^{-1/2}
\end{equation}
\citep[see, e.g., eqs.~(17.2.8) and~(17.3.5) in][]{ShaT83};
here the coefficient has been evaluated for a composition of pure
ionized hydrogen.  For SMSs with $M \gtrsim 10^4 M_{\odot}$, we can
therefore neglect the pressure contributions of the plasma in
determining the equilibrium profile, even though the plasma may be
important for determining the stability of the star \citep{ZelN71,ShaT83}. 
A simple proof that
very massive stars or SMSs are convective in this limit has been given by 
\cite{LoeR94}.\footnote{In fully convective stars energy transport in the stellar
interior is dominated by convection. However, there is still an outgoing flux of radiation from the stellar surface.  The later is what we compute in Section 4 below.}
This result implies that the photon entropy per baryon, 
\begin{equation}
s_r = \frac{4}{3} \,\frac{a T^3}{n_B} 
\end{equation}
is constant throughout the star, and so therefore is $\beta_P \approx 8
(s_r/k)^{-1}$.  Here $a$ is the radiation density constant, $T$ is the temperature, $n_B$ is the
baryon density, and $k$ is Boltzmann's constant.  As a consequence,
the equation of state of a very massive star or SMS is that of an $n=3$ polytrope:
\begin{equation}
P = K \rho^{4/3},  \mbox{~~}
K = \left[ \left( \frac{k}{\mu m} \right)^4 \frac{3}{a}
\frac{(1 + \beta_P)^3}{\beta_P^4} \right]^{1/3} = {\rm const},
\end{equation}
where $m$ is the atomic mass unit and $\mu$ is the mean molecular
weight \citep[cf.][eq.~(2-289); note that Clayton adopts a
different definition of $\beta_P$, which is related to ours by
$\beta_{\rm Clayton} = \beta_P /(1+\beta_P)$]{Cla83}.  In the 
high temperature, low density, strongly ionized plasma of a
very massive or SMS, Thomson scattering off free electrons is the dominant source of
opacity.  This opacity is independent of density and justifies our
assumption about the Rosseland mean opacity.

In applications to SMSs our analysis neglects electron-positron
pairs and Klein-Nishina corrections to the
electron-scattering opacity, which is valid for $M \ga10^5 M_{\odot}$
\citep[see, e.g.][]{FulWW86}.

\section{The Roche model for a binary}
\label{sec:roche}

We begin with the equation of hydrostatic equilibrium satisfied by the primary,
\begin{equation} \label{hydrostatic_eq}
\frac{\nabla P}{\rho} = - \nabla ( \Phi_p + \Phi_c +\Phi_r).
\end{equation}
Here the right-hand side describes an effective gravitational force, which is derived from the
(interior)  gravitational potential $\Phi_p$ of the primary, the (exterior) gravitational potential of the companion $\Phi_c$, and 
a centrifugal potential $\Phi_r$ arising from the (synchronous) rotation of the primary.

Stars with soft equations of state are centrally condensed, i.e.~most of the mass is concentrated in a high-density core that is 
surrounded by an extended  low-density envelope. For an $n=3$ polytrope, for example, the ratio between central and mean density 
is $\rho_c/\bar \rho = 54.2$.  The gravitational force in the envelope is therefore dominated by the massive core, and it is thus legitimate to neglect the self-gravity of the envelope.   In the envelope, we may therefore approximate the Newtonian potential of the primary star $\Phi_p$ as
\begin{equation}
\Phi_p = -\frac{M_p}{r}
\end{equation}
where $M_p$ is the mass of the primary and $r$ the distance from the primary's center (here we adopt gravitational units by setting $G \equiv 1$).  

\begin{figure}
\includegraphics[width=3in]{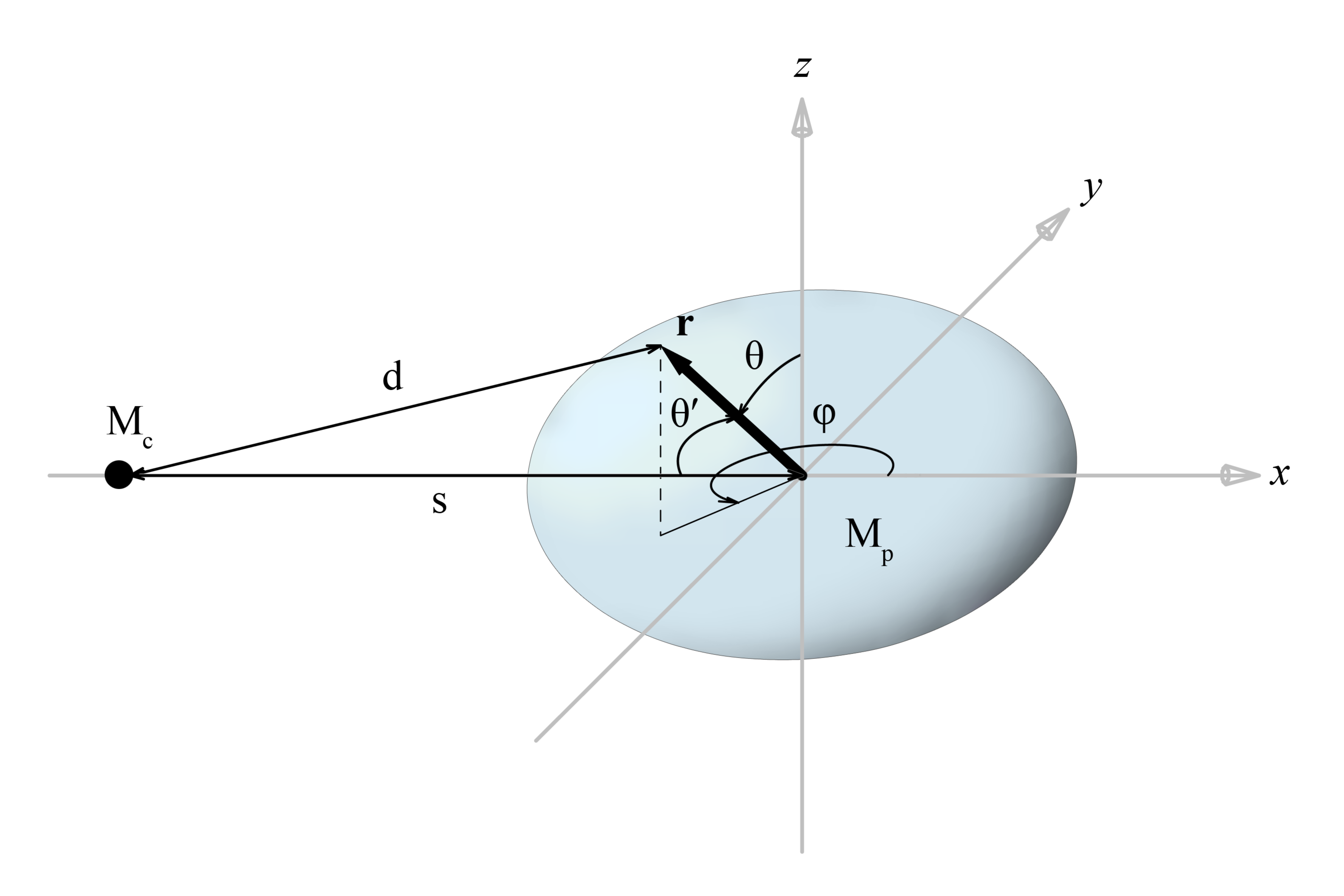}
\caption{Sketch of the coordinate system used in our calculation.  The primary of mass $M_p$ is represented by the shaded configuration on the right,  while the companion of mass $M_c$ is represented by the dot on the left.\\ \label{Fig1}}
\end{figure}

In accord with the Roche model, the companion may be treated as a point mass $M_c$, located at a distance $s$ from the center of the primary (see Fig.~\ref{Fig1}).  The potential at a point ${\bf r}$ in the primary is then given by 
\begin{equation}
\Phi_c = - \frac{M_c}{d},
\end{equation}
where $d$ is the distance from $M_c$.  We now expand $d$ about $s \geq r$,
\begin{equation} \label{expand}
\frac{1}{d} = \frac{1}{s} \sum_{\ell = 0}^\infty \left( \frac{r}{s} \right)^\ell P_\ell(\cos \theta'),
\end{equation} 
where $P_\ell(\cos \theta')$ is the Legendre polynomial of order $\ell$, and where $\theta'$ is the angle between ${\bf r}$ and 
the line connecting the center of the primary to the point mass companion. The first term  $\ell = 0$ is a constant that can be ignored.   We will see that the $\ell = 1$ term will cancel out later when we consider the rotational contribution $\Phi_r$,  
but we retain it for now. The first relevant term is the quadrupolar tidal term $\ell = 2$.  Truncating the expansion (\ref{expand}) after this term we approximate
\begin{equation}
\Phi_c = - \frac{M_c r}{s^2} \cos \theta' - \frac{M_c r^2}{2 s^3} \left( 3 \cos^2 \theta' - 1 \right).
\end{equation}
We now introduce a coordinate system as shown in Fig.~\ref{Fig1}, so that the orbital plane is in the $x-y$ plane, with the center 
of the primary at the origin and the companion at $x = -s$ and $y=z=0$.  In terms of spherical polar coordinates we now express $\cos \theta' = - x/r = - \cos \phi \sin \theta$, and hence
\begin{equation} \label{Phi_c}
\Phi_c = \frac{M_c x}{s^2} - \frac{M_c}{2} \frac{r^2}{s^3} (3\cos^2 \phi \sin^2 \theta - 1).
\end{equation}

Finally, the rotational potential $\Phi_r$ in (\ref{hydrostatic_eq}) arises from the rotation of the primary about the system's center of mass.  Assuming corotation, so that the star appears static in the rotating frame of the binary, we may write this term as 
\begin{equation} \label{Phi_r}
\Phi_r = -\frac{1}{2} \Omega^2 \left( (x - x_{\rm CM})^2 + y^2 \right),
\end{equation}
where 
\begin{equation} \label{CM}
x_{\rm CM} = - \frac{M_c}{M_p + M_c} s
\end{equation}
is the location of the center of mass.   Using (\ref{CM}) as well as Kepler's law 
\begin{equation} \label{Kepler}
\Omega^2 = \frac{M_p + M_c}{s^3}
\end{equation}
we can write (\ref{Phi_r}) as 
\begin{equation} \label{Phi_r_2}
\Phi_r = - \frac{1}{2} \Omega x_{\rm CM}^2 - \frac{M_c x}{s^2} - \frac{1}{2} \frac{M_p + M_c}{s^3} r^2 \sin^2 \theta.
\end{equation}
The first term on the right-hand side is a constant term that can be ignored.   The second term will exactly cancel the first term in (\ref{Phi_c}) when the two potentials are added in (\ref{hydrostatic_eq}); we may therefore discard the linear terms in both (\ref{Phi_c}) and (\ref{Phi_r_2}).  Our potentials $\Phi_c$ and $\Phi_r$ then reduce to
\begin{eqnarray} \label{Phi_c_f}
\Phi_c  + \Phi_r &=&  - \frac{M_c}{2} \frac{r^2}{s^3} (3\cos^2 \phi \sin^2 \theta - 1) \nonumber \\  
                & &  - \frac{1}{2} \frac{M_p + M_c}{s^3} r^2 \sin^2 \theta. 
\end{eqnarray}

Integrating eq.~(\ref{hydrostatic_eq}) yields the Euler integral
\begin{equation} \label{Euler}
h + \Phi_p +\Phi_c +\Phi_r = H,
\end{equation}
where H is a constant of integration and where 
\begin{equation}
h = \int{\frac{dP}{\rho}} = (n+1)\frac{P}{\rho}
\end{equation}
is the enthalpy per unit mass.  Evaluating (\ref{Euler}) for infinite binary separation $s \rightarrow \infty$ (where $\Phi_c = \Phi_r = 0$) at the stellar surface (where $h = 0$) we find
\begin{equation} \label{H}
H = - \frac{M_p}{R_0},
\end{equation} 
where $R_0$ is the stellar radius for the nonrotating (spherical) star in isolation.   

It is consistent with the Roche approximation to assume that the core remains unaffected by the presence of the binary companion.   Given that $\Phi_c$ and $\Phi_r$ vanish at the center of the star, and given that we may assume $h$ and $\Phi_p$ to remain 
unperturbed there, we may also assume that $H$ is independent of the binary separation, so that we can always express it in terms of $R_0$ as in eq.~(\ref{H}).  Similar arguments have been made for isolated rotating stars \cite[e.g.][]{ZelN71,ShaT83}, where
they have been confirmed by numerical simulations \cite[e.g.][]{PapW73}.

\begin{figure}
\includegraphics[width=3in]{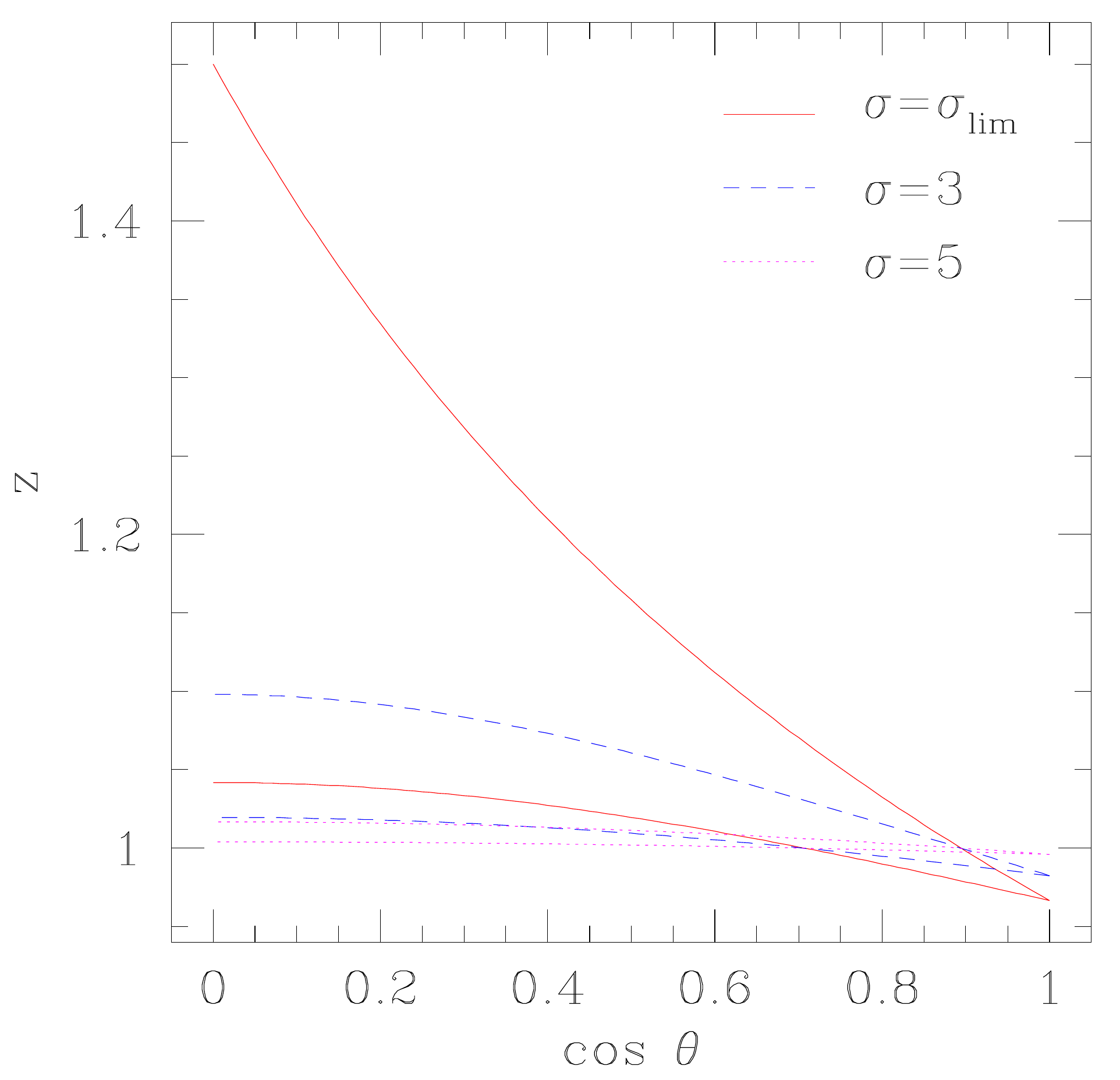}
\caption{The dimensionless stellar radius $z$ 
as a function of $\cos \theta$ for an equal-mass binary ($q=1$) at different values of the binary separation $\sigma$.  For each binary separation the top line represents results for $\phi = \pi$ (i.e.~in the direction toward the binary companion) while the bottom line represents results 
for $\phi = \pi/2$.  The two lines connect at the pole ($\cos \theta = 1$).  
Note that for $\sigma = \sigma_{\rm lim}$, we have $z = 3/2$ for the 
point facing the binary companion (see eq.~(\ref{z_lim})). 
\label{Fig2}}
\end{figure}

The surface of the star satisfies $h=0$ and can be determined from eq.~(\ref{Euler}) in the form 
\begin{equation}  \label{surface_1}
\Phi_p + \Phi_c + \Phi_r - H = 0.
\end{equation}
Introducing dimensional parameters for the mass ratio
\begin{equation}
q \equiv  \frac{M_c}{M_p},
\end{equation}
the binary separation
\begin{equation}
\sigma \equiv \frac{s}{R_0},
\end{equation}
and the distance from the stellar center to the surface
\begin{equation}
z \equiv \frac{r}{R_0},
\end{equation}
we can bring (\ref{surface_1}) into the form of a cubic equation for $z$,
\begin{equation} \label{surface_2}
C_3 z^3 - z + 1 = 0,
\end{equation}
where we have abbreviated
\begin{equation} \label{C3}
C_3 = \frac{1}{2\sigma^3} \Big( q (3\cos^2\phi \sin^2\theta - 1) + (1+q) \sin^2\theta \Big). 
\end{equation}
Given binary parameters $q$ and $\sigma$, this cubic equation can be solved analytically to find $z$ as a function of the coordinates $\theta$ and $\phi$ (see Appendix).

Note that the cubic (\ref{surface_2}) takes exactly the same form as that for a single rotating star (see, eq.~(23) in Paper I), except that the coefficient $C_3$ is now different.\footnote{Our coefficient $C_3$ reduces to the corresponding coefficient for rotating stars when $q = 0$.}  Instead of depending on $\theta$ only, as for axisymmetric, rotating stars, it now depends on both $\theta$ and $\phi$.   Moreover, for rotating stars the corresponding coefficient is nonnegative, resulting in values for $z$ that are always greater than or equal to unity.  Here, however, $C_3$ can be positive or negative.  At the poles, where $\sin \theta = 0$, for example, we have $C_3 < 0$ (for finite $\sigma$), resulting in $z < 1$.  This is consistent with the fact that the tidal field of the companion leads to the squeezing of the primary along its poles and an elongation along its equator.  

Hydrostatic equilibrium is only possible if, at the surface of the star, where the density vanishes, the pressure increases towards the interior of the star.  Our sequences of hydrostatic equilibria therefore terminate when, at the point facing the companion, i.e.~at $\theta = \pi/2$ and $\phi = \pi$, the right-hand side of (\ref{hydrostatic_eq}) vanishes.\footnote{Having truncated the expansion (\ref{expand}) after the tidal term, this point is equivalent to the point pointing away from the binary companion, $\theta = \pi/2$ and $\phi = 0$.}   Evaluating the right-hand side of (\ref{hydrostatic_eq}) at that point, and setting it to zero, yields
\begin{equation} \label{z3_limit}
(1 + 3 q) \frac{z^3}{\sigma^3} = 1.
\end{equation}
For $\theta = \pi/2$ and $\phi = \pi$ we also have
\begin{equation} \label{C3_limit}
C_3 = \frac{1 + 3q}{2\sigma^3}. 
\end{equation}
Inserting (\ref{z3_limit}) together with (\ref{C3_limit}) into the cubic (\ref{surface_2}) we find that the limiting value of $z$, at $\theta = \pi/2$ and $\phi = \pi$, is
\begin{equation} \label{z_lim}
z_{\rm lim} = \frac{3}{2}.
\end{equation}
Interestingly, this is the same value obtained for the equator of single stars rotating at the break-up limit (see, e.g., eq.~(10) in Paper I).  Inserting (\ref{z_lim}) back into (\ref{z3_limit}) we now obtain the limiting binary separation $\sigma_{\rm lim}$ for which, under our assumptions, sequences of hydrostatic equilibria end,
\begin{equation} \label{sigma_lim}
\sigma_{\rm lim} = \frac{3}{2} (1 + 3q)^{1/3}.
\end{equation}
For a mass ratio of $q = 1$, for example, we have 
$\sigma_{\rm lim} \approx 2.38$. 
Given that $\sigma \ga 2.4$ for all equilibrium models, our truncation of the
interaction potential beyond the tidal term is justified in a 
first approximation.
The corresponding Roche limit for a homogeneous, incompressible ($n=0$) star, allowing for departure of the angular velocity from the Keplerian value due to the ellipsoidal shape of the primary, yields $\sigma_{\rm lim} = 2.713$ \citep{LaiRS93}.

In Fig.~\ref{Fig2} we show results for $z$ for an equal-mass binary for different values of the binary separation.  In particular, these results confirm the limiting value (\ref{z_lim}) for the end point of the sequence at $\sigma_{\rm lim}$.

\section{Gravity darkening and brightening}
\label{sec:flux}

\begin{figure}
\includegraphics[width=3in]{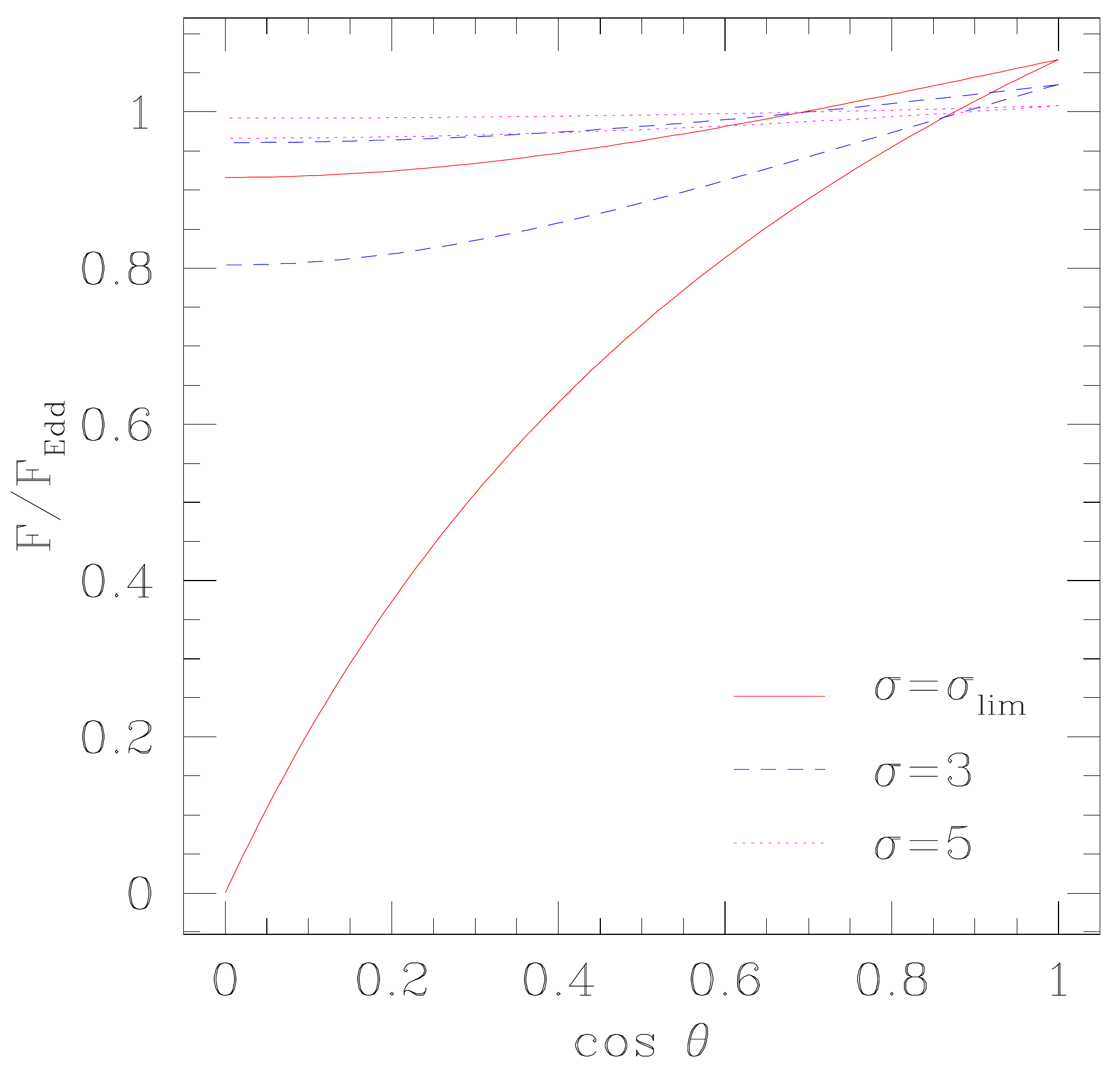}
\caption{The flux $F/F_{\rm Edd}$ as a function of function of $\cos \theta$ for an equal-mass binary ($q = 1$) at different values of the binary separation $\sigma$.  For each binary separation the bottom line represents results for $\phi = \pi$ (i.e.~in the direction toward the binary companion) while the top line represents results for $\phi = \pi/2$.  The two lines connect at the pole ($\cos \theta = 1$).   Note that in a region around the pole the flux exceeds the corresponding Eddington flux, and that the flux vanishes at the point facing the binary companion when $\sigma = \sigma_{\rm lim}$.  \label{Fig3}}
\end{figure}

In the diffusion approximation, the radiation flux is given by
\begin{equation} \label{flux1}
\mathbf{F} = -\frac{1}{3\kappa \rho}\mathbf{\nabla} U,
\end{equation}
where $U$ is the energy density of the radiation,
\begin{equation} \label{U}
U = aT^4 = 3P,
\end{equation} 
and where $\kappa$ is the opacity (which we assume to be dominated by electron scattering, $\kappa = \kappa_{\rm es}$).  We have also assumed that the pressure is dominated by radiation pressure
\begin{equation} 
P \approx P_r = \frac{1}{3}aT^4.
\end{equation}

Inserting equations (\ref{flux1}) and (\ref{U}) into the equation of hydrostatic equilibrium (\ref{hydrostatic_eq}) yields
\begin{equation}
\kappa \mathbf{F}= \mathbf{\nabla}(\Phi_p + \Phi_c + \Phi_r).
\end{equation}
In polar coordinates in an orthonormal basis, the magnitude of the flux is
\begin{equation}
F = (F^2_{\hat{r}} +F^2_{\hat{\theta}} +F^2_{\hat{\phi}})^{1/2}.
\end{equation}
Evaluating the gradients of the potentials $\Phi_p$, $\Phi_c$ and $\Phi_r$ we find
\begin{eqnarray} \label{flux}
\frac{F}{F_{\rm Edd}} & = & \Big\{\left(1- \frac{q}{\sigma^3}z^3 (3\cos^2\phi \sin^2\theta - 1) 
- \frac{1+q}{\sigma^3} z^3 \sin^2\theta\right)^2  \nonumber \\
& & + \left(\frac{q}{\sigma^3}z^3(3\cos^2\phi \cos\theta \sin\theta) + \frac{1+q}{\sigma^3} z^3 \sin\theta \cos\theta\right)^2 \nonumber \\
& & 
+ \left(\frac{q}{\sigma^3}z^3(3\cos\phi \sin\phi \sin\theta) \right)^2 \Big\}^{1/2} 
\end{eqnarray}
where 
\begin{equation}
F_{\rm Edd} = \frac{M_p}{\kappa r^2}
\end{equation}
is the Eddington flux from a spherical star of radius $r$. 

\begin{figure}
\begin{center}
\framebox{\includegraphics[width=2in]{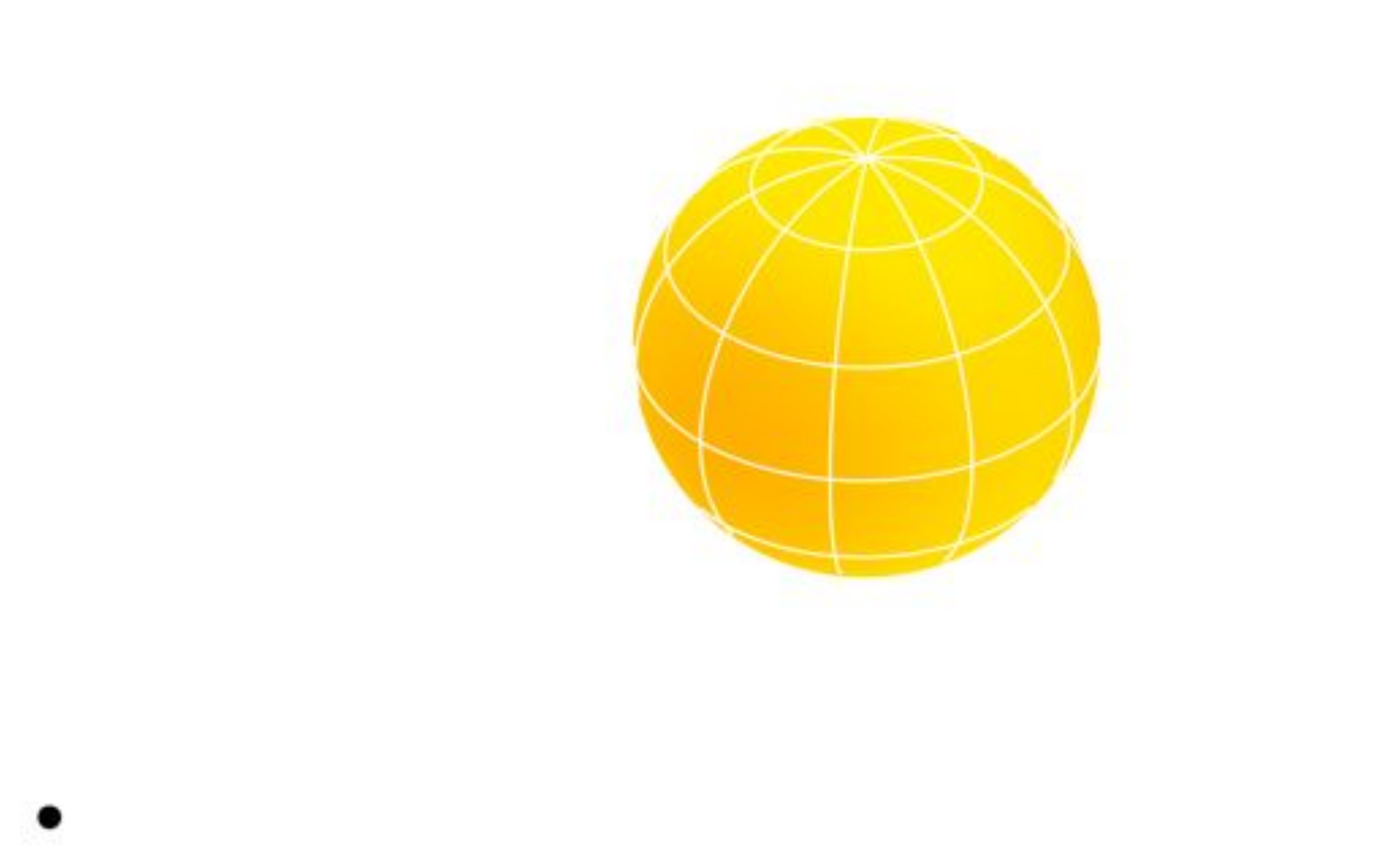}}
\framebox{\includegraphics[width=2in]{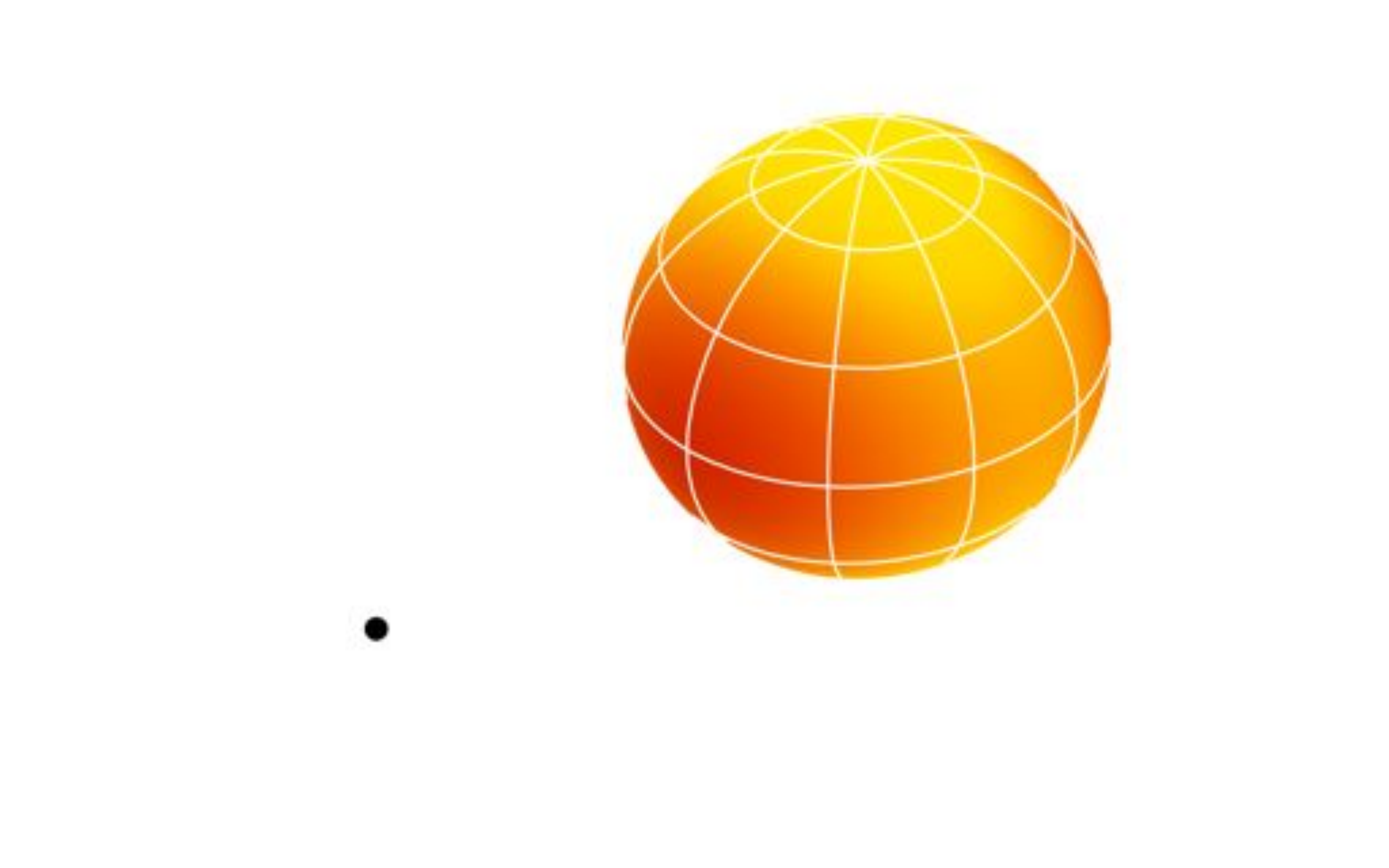}}
\framebox{\includegraphics[width=2in]{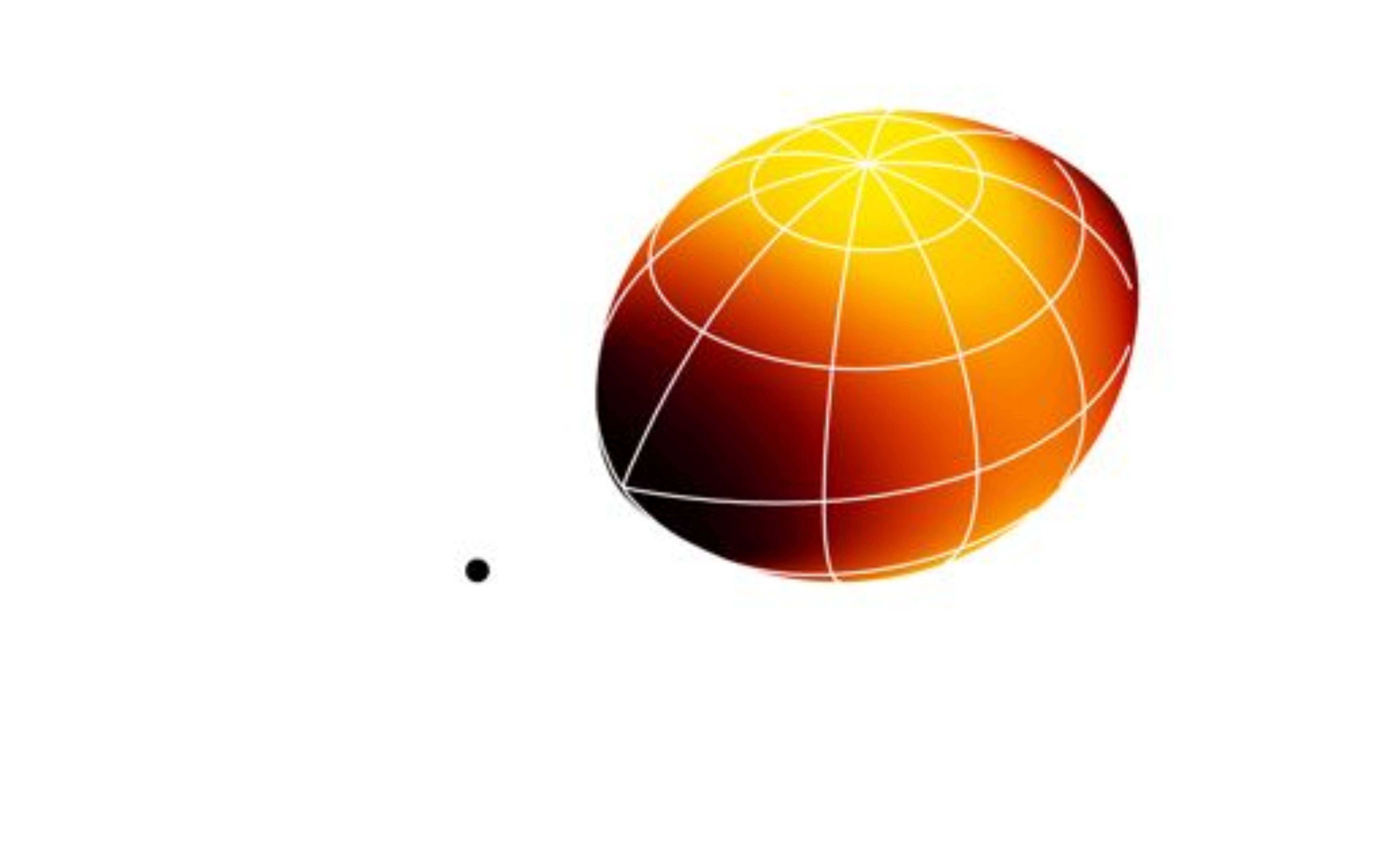}}
\end{center}
\caption{Surface images of the primary in an equal-mass binary at three different binary separations $\sigma = 5$ (top panel), $\sigma = 3$ (middle panel) and $\sigma = \sigma_{\rm lim}$ (bottom panel).  The shape of the primary is given by eq.~(\ref{surface_2}), while the color coding represents the flux (\ref{flux}).  Yellow indicates the largest flux, red a smaller flux, and black a vanishing flux.  The dot represents the point mass that models the companion.
\label{Fig4}}
\end{figure}

We graph the flux $F$ for an equal-mass binary for different values of the binary separation $\sigma = s/R_0$ in Fig.~\ref{Fig3}.  In a region around the poles the flux is greater than the Eddington flux.  The reason for this ``gravity brightening" effect is that, at the poles, the tidal forces caused by the companion lead to an increase of the effective gravitational force, which in turn leads to an increase in the flux.  The opposite is true at the points on the stellar equator either pointing directly toward the companion or directly away from the companion.   At these points, the tidal forces lead to a reduction in the effective gravitational force.  On the equator, stellar rotation also leads to a reduction of the effective gravitational force.  At the two points facing towards or away from the companion (i.e.~for $\phi = \pi$ or $\phi = 0$), the two effects act together to result in the greatest reduction in the brightness, i.e.~the strongest gravity darkening effect.  For the limiting binary separation $\sigma = \sigma_{\rm lim}$, the flux completely vanishes at those points.  At the two points on the equator with $\phi = \pi/2$ and $\phi = 3 \pi /2$, on the other hand, the two effects counteract, leading to a smaller reduction in the flux. 

In Fig.~\ref{Fig4} we also show surface images of the primary in an equal-mass binary at three different binary separations.  The shape of the primary is given by eq.~(\ref{surface_2}) and reflects the tidal deformation, while the color coding indicates the local radiative surface flux (\ref{flux}).  Yellow indicates the largest flux, red a smaller flux, and black a vanishing flux.

\section{Luminosity}
\label{sec:luminosity}

\begin{figure}
\includegraphics[width=3in]{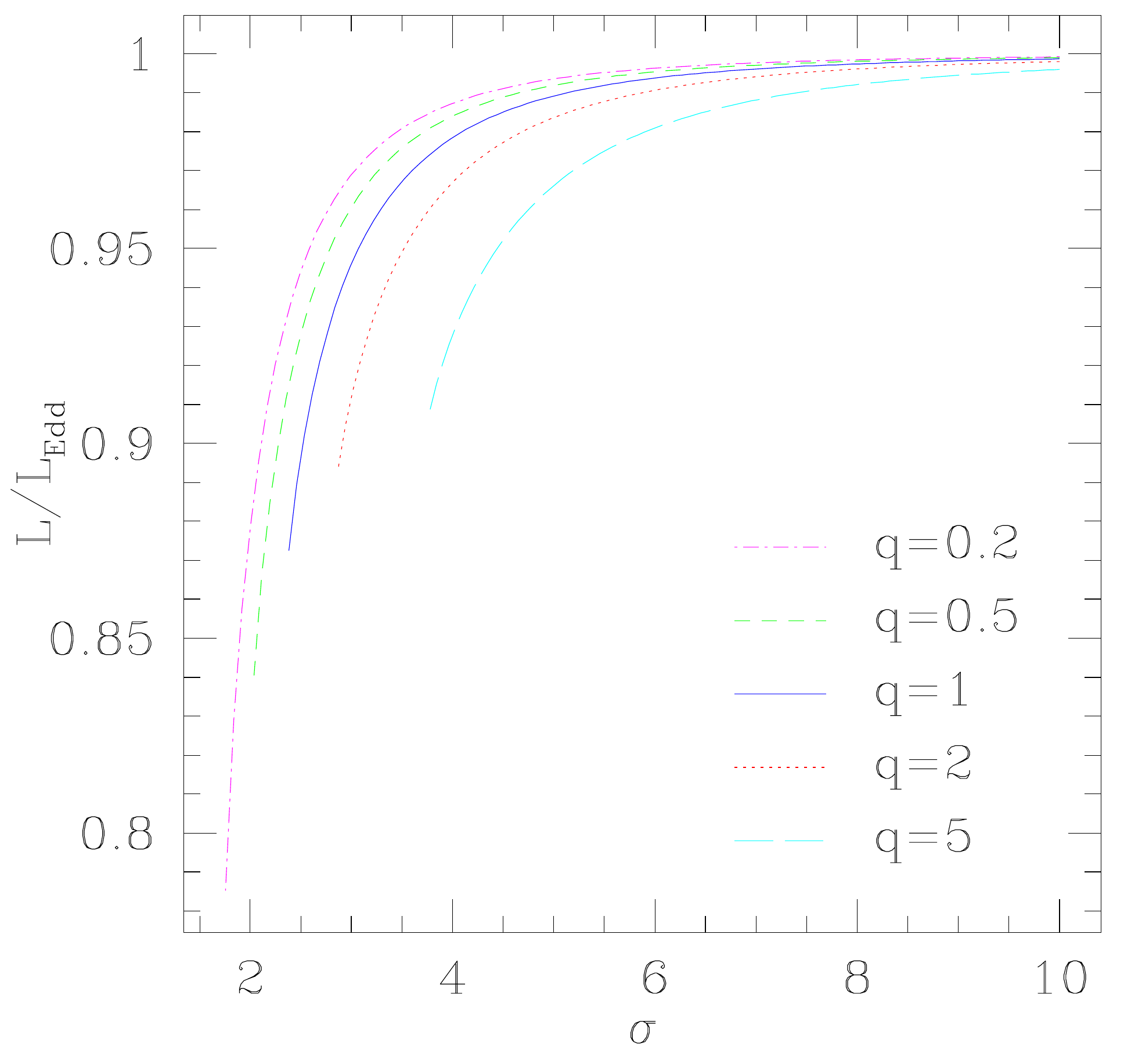}
\caption{The luminosity $L/L_{\rm Edd}$ as a function of binary separation $\sigma$ for different values of the mass ratio $q$.  For each value of $q$, the sequences start at $\sigma = 10$ and end at $\sigma_{\rm lim}$ given by eq.~(\ref{sigma_lim}). \label{Fig5}}
\end{figure}

We find the total luminosity of the star by integrating the flux (\ref{flux}) over the stellar surface
\begin{equation}
L = \oint {\bf F} \cdot d {\bf {\mathcal A}} = \oint F d {\mathcal A}.
\end{equation}
The surface element can be written as
\begin{equation}
d {\mathcal A} = \left\{ 1 + \frac{1}{z^2} \left( \frac{dz}{d\theta} \right)^2 + 
\frac{1}{z^2\sin^2 \theta} \left( \frac{dz}{d\phi} \right)^2 \right\}^{1/2} r^2 d \Omega,
\end{equation}
where $d \Omega = \sin \theta d \theta d \phi$.   We then insert the flux (\ref{flux}) and evaluate the integration numerically to find the star's luminosity $L$.  It is convenient to divide the result by the Eddington luminosity
\begin{equation}
L_{\rm Edd} = \frac{4 \pi M}{\kappa}
\end{equation}
and express the results in terms of the dimensionless ratio $L/L_{\rm Edd}$.  In Fig.~\ref{Fig5} we show results for the luminosity as a function of the binary separation $\sigma$ for different values of the mass ratio $q$.  All
sequences shown start at $\sigma = 10$ and end at the limiting binary separation $\sigma_{\rm lim}$ given by eq.~(\ref{sigma_lim}).  For an equal-mass binary, for example, the luminosity is reduced by about 12.7\% when the binary reaches $\sigma_{\rm lim}$.  We also verified that in the limit $q \rightarrow 0$, the luminosity for $\sigma = \sigma_{\rm lim}$ is reduced by about 36\%, which is the value found for single stars rotating at the break-up limit (see, e.g., Paper I).

\section{Discussion}
\label{sec:discussion}

We apply gravity darkening models to corotating binary stars and obtain simple and analytical expressions for the surface flux of tidally distorted stars.  The tidal interaction leads to both gravity darkening (along the equator) and gravity brightening (in regions around the poles).  We identify a critical separation at which, within the Roche model, sequences of hydrostatic equilibrium end, and at which the radiative flux at the point on the equator that faces the binary companion vanishes.  At this critical separation, the total luminosity in an equal-mass binary is about 13\% less than for the corresponding nonrotating star in isolation.

Simple and analytical models for the flux and luminosity from binary
stars, even if they are approximate, are useful in many ways.  In
particular, we hope that they will provide useful models for
comparison with future resolved interferometric images of close binary
stars.  For single rotating stars, von Zeipel gravity darkening models
capture the basic features of interferometric images reasonably well,
but they also show some deviations.  Presumably, these deviations are
caused by the fact that some of the assumptions do not apply to the
observed stars.  In particular, many main sequence stars have
complicated surface layers or rotate differentially rather than
uniformly.  As discussed by \cite{Monetal07} \citep[see
also][]{Zhaetal09}, the agreement between model and observations can
be improved by introducing new free parameters.  \cite{Monetal07} and
others have used ``$\beta$-free" models, in which the effective
temperature is taken to be $T_{\rm eff} \propto g_{\rm eff}^\beta$,
where $g_{\rm eff}$ is the effective gravitational force on the
right-hand side of eq.~(\ref{hydrostatic_eq}).  In the ``standard"
model, $\beta = 1/4$, but allowing $\beta$ to be a free parameter
allows for improved fits to the observations.  Other improvements to
the simple von Zeipel models have been proposed by \cite{EspR11} and
\cite{Cla12}.  We expect that similar generalizations may improve
future fits between binary gravity darkening models and observations
as well.

\acknowledgments

It is a pleasure to thank Andrew Currier for producing Figs.~1 and 4 for us.  HEW gratefully acknowledges support through a Clare Boothe Luce undergraduate fellowship.  This work was supported in part by NSF Grant PHY-1063240 to
Bowdoin College and by NSF Grant PHY-0963136 as well as 
NASA Grant NNX10AI73G at the University of Illinois at Urbana-Champaign.

\appendix
\section{Solution for the stellar surface}

In this brief appendix we present the solution to the cubic equation (\ref{surface_2}), yielding the stellar surface $z$.  The general solution to a cubic equation can be found, for example, in \cite{PreTVF07}.  Applying their prescription to our equation (\ref{surface_2}), we see that the form of the solution $z$ depends on the coefficient $C_3$.  

By combining equations (\ref{C3}) and (\ref{sigma_lim}) we first observe that we always have $C_3 
\leq 4/27$.  The form of the solution then depends only on the sign of $C_3$.  If $C_3$ is positive, then the cubic has three real roots and we pick the one that yields $z = 1$ when $C_3 = 0$, given by
\begin{equation}
z = - \frac{2}{\sqrt{3 C_3}} \cos \left( \frac{ \arccos( \sqrt{27 C_3/4}) - 2 \pi}{3} \right)~~~~~~~~~~(C_3 \geq 0).
\end{equation}
For a single, rotating star, we always have $0 \leq C_3 \leq 4/27$ (see, e.g., Paper I), so that the solution can always be written in this form.  For a binary, however, $C_3$ can also be negative.  In this case, the cubic has one real and two imaginary roots.  Defining
\begin{equation}
A = \left( \frac{1}{2 |C_3|} \left\{ \left(1 + \frac{4}{27 |C_3|} \right)^{1/2} + 1 \right\} \right)^{1/3}
\end{equation}
we can write the real root as
\begin{equation}
z = A - \frac{1}{3 |C_3| A}~~~~~~~~~~(C_3 < 0).
\end{equation}

\end{document}